\documentclass[10pt]{article}
\usepackage{geometry} 
\usepackage{color}
\usepackage{graphicx}
\pdfoutput=1
\title{Tradition versus fashion in consumer choice}
\author{R. Alexander Bentley\\Anthropology Department, Durham University \\Durham DH1 3LE UK\\r.a.bentley@durham.ac.uk \and Paul Ormerod\\Institute of Advanced Study, Durham University \\ Volterra Consulting Ltd. \\ London SW14 8AE, UK\\pormerod@volterra.co.uk}
\begin{document}
\maketitle
\begin{abstract}
Evidence is growing that in many markets consumers select not simply on the basis of the perceived attributes of products,  but their preferences are modified by the behaviour of others.  Economists have paid relatively little attention  to such  markets. The classic Bass diffusion model incorporates the imitation of others as a part of the behavioural rules used by consumers in making choice. We extend the Bass model to be able to apply it to long-term case studies where substantial changes over time in the population making choices at any given point have to be taken into account. We consider evidence from the activity of hill-walking.  We use as a particular illustration the Munros, a list of Scottish hills over 3,000 feet in height which many British walkers aspire to complete.  The extension to the Bass model can of course be used in other situations where such population changes are important.
\end{abstract}
%
%
\section{Introduction}
Standard consumer choice theory assumes atomised individuals exercise choice in an attempt to maximise utility subject to a budget constraint.  In this approach, given an individual's tastes and preferences, decisions are taken on the basis of the attributes of the various products, such as price and quality.  
\\ \indent In recent decades, the conventional theory has been extended to allow for factors such as the cost of gathering information \cite{Stigler_1961}, imperfections in the perception of information and limitations to consumers' cognitive powers in gathering and processing information \cite{Simon_1955}.  
\\ \indent In general, however, economists have paid little attention to markets in which fashion is important \cite{Chai_etal_2007}; i.e., markets in which the decisions of others can affect directly the choices made by an individual. Interest in such markets has, however, been much greater outside of economics.  For example, two recent papers by non-economists \cite{Salganik_etal_2006, Colbaugh_Glass_2007} provide strong empirical evidence that in markets where the decisions by others strongly influence individual choice, products which are superior in terms of their attributes may do no better than ones which are worse.
\\ \indent We set out an extension to the existing analysis of markets where consumers do not select products (or behaviours) solely on the basis of inherent attributes, but also imitate what other people are doing. 
\\ \indent Our extension is to long-term case studies where substantial changes over time in the population making choices at any given point have to be taken into account.  We may usefully posit two potential extremes of a spectrum of behaviour in such circumstances. At one end, we have what we might label a `tradition,' which is carried on by new generations, and at the other, a `fashion', which has most of its impact on one particular generation. 
\\ \indent We illustrate the approach using data maintained by the Scottish Mountaineering Council (SMC), who keep a list of people who have completed the `Munros', a specific list of 284 hills in Scotland with a height of at least 3,000 feet. 
\section{The data}
Despite the relatively low height of the summits, the Munros are nevertheless a demanding group of hills, as the summits are fully exposed to high winds, rain and mist, and on the highest hills snow can fall on any day of the year.  But they are an attractive challenge within the capability of fit walkers willing to dedicate time and effort. Substantial numbers of walkers aspire to complete this list,  whereupon they can register a claim with the SMC.
\\ \indent The actual number of Munros completions registered each year (Figure 1) show that it peaked in 1999 at 247, and appears to have levelled off, averaging\footnote[1]{The  data were taken from the SMC website in early March 2008. There are a number of issues around the accuracy of the data, which are dealt with in a longer version of the paper forthcoming in the \textit{Scottish Journal of Political Economy}, available on request from the authors.} 203 between 2000 and 2007. 
%
%
\begin{figure}
\begin{center}
\includegraphics[width=5.5in]{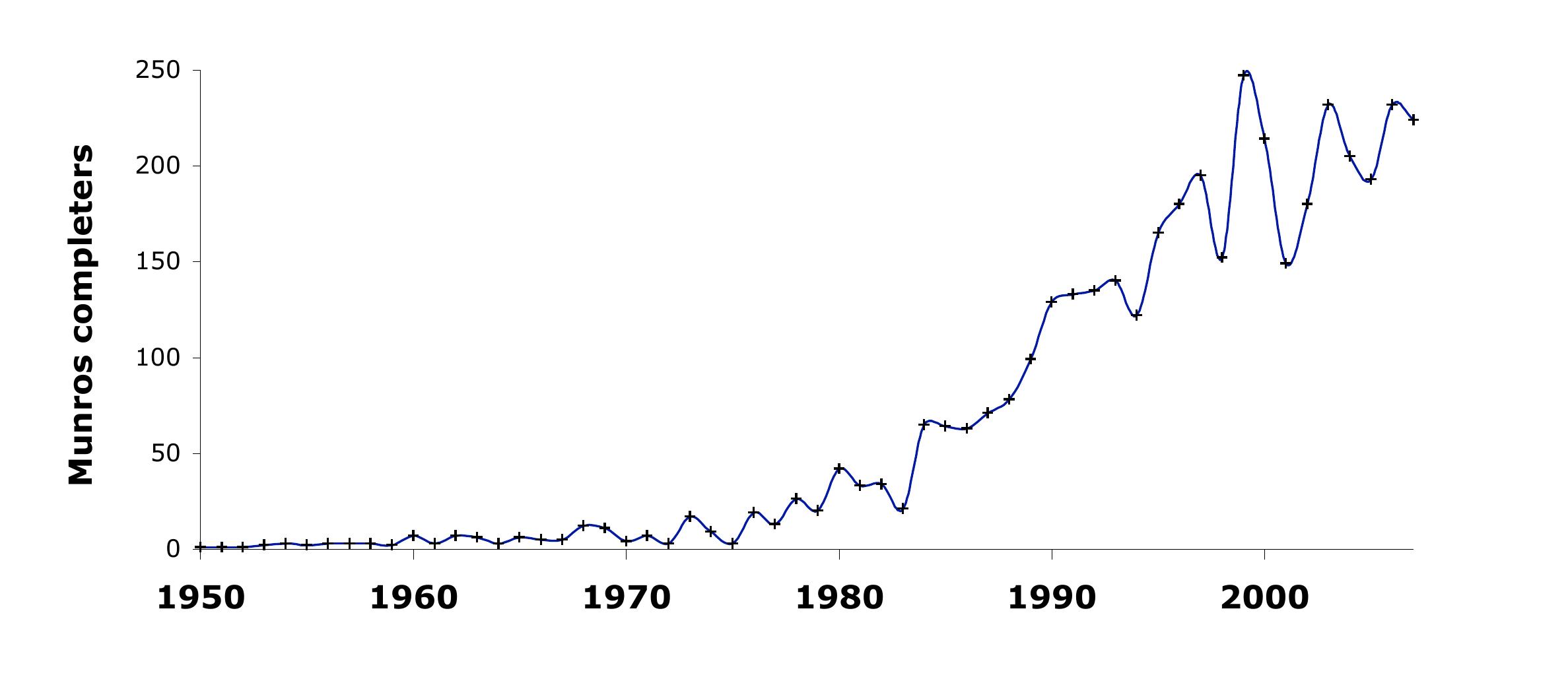}
\end{center}
\caption{Number completing the list of Scottish Munros each year, by year of completion.  Source: Scottish Mountaineering Council, www.smc.org.uk.}
\end{figure}
\\ \indent The Munros are not without their own variant lists.  The principal one is the Munro Tops, which are hills in Scotland which are also above 3000 feet but which for some reason are not regarded as separate mountains. A fuller discussion of this subtlety is in our forthcoming \textit{Scottish Journal of Political Economy} article. The total number of Munro completers is 4019, but the number of Top completers is only 446.  
\\ \indent The annual completions of the Tops rose sharply during the 1980s, but in contrast to the Munros, its popularity has declined sharply of late, as the annual number of completions shows (Figure 2a). Figure 2b shows this decline is of gradual exponential form when the Top completions are expressed as a fraction of those completing the Munros (there were only 17 Top completions in total before 1965, so Figure 2 plots the data from 1965 onwards).
%
%
\begin{figure}
\begin{center}
\includegraphics[width=2.8in]{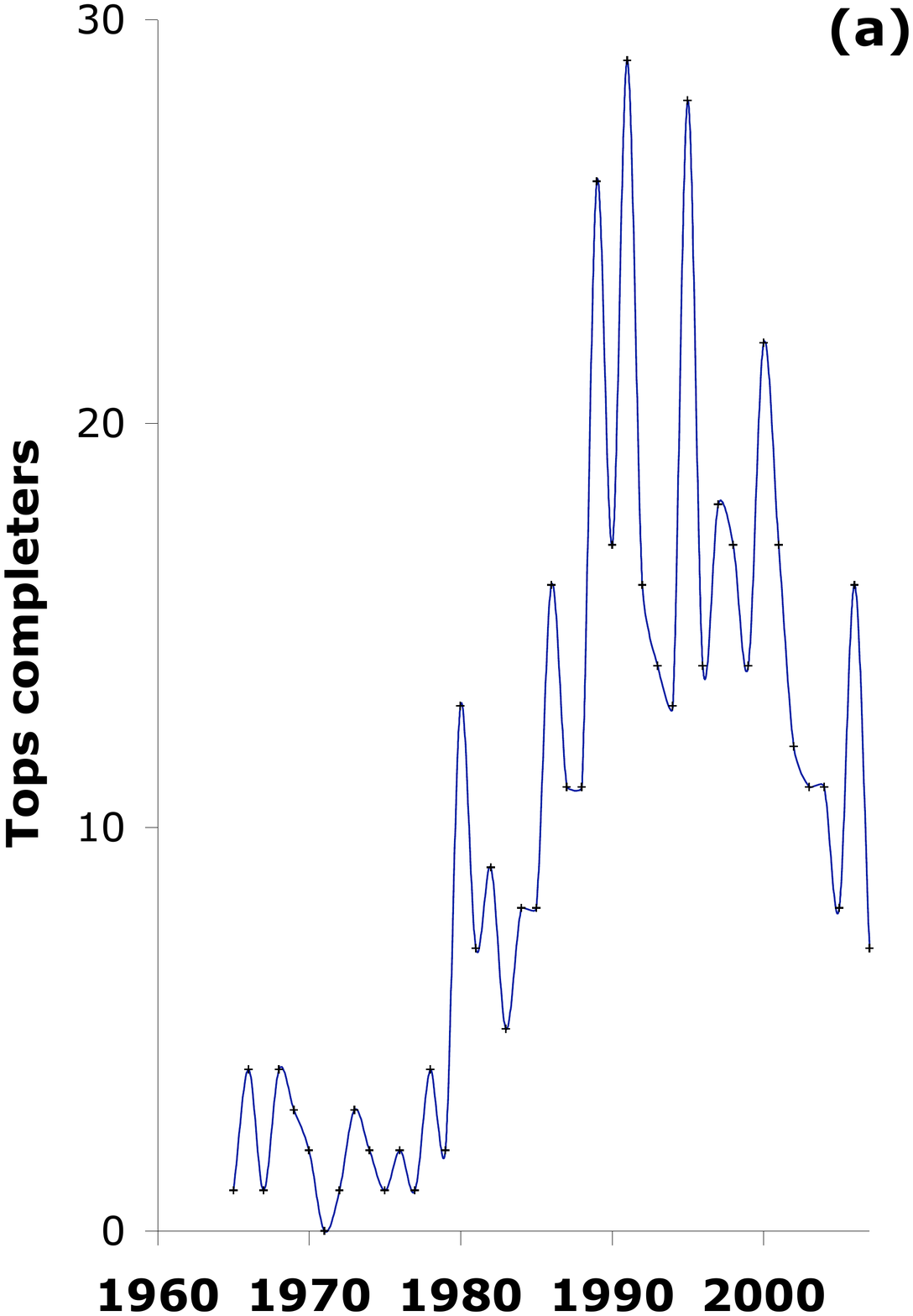}
\includegraphics[width=2.8in]{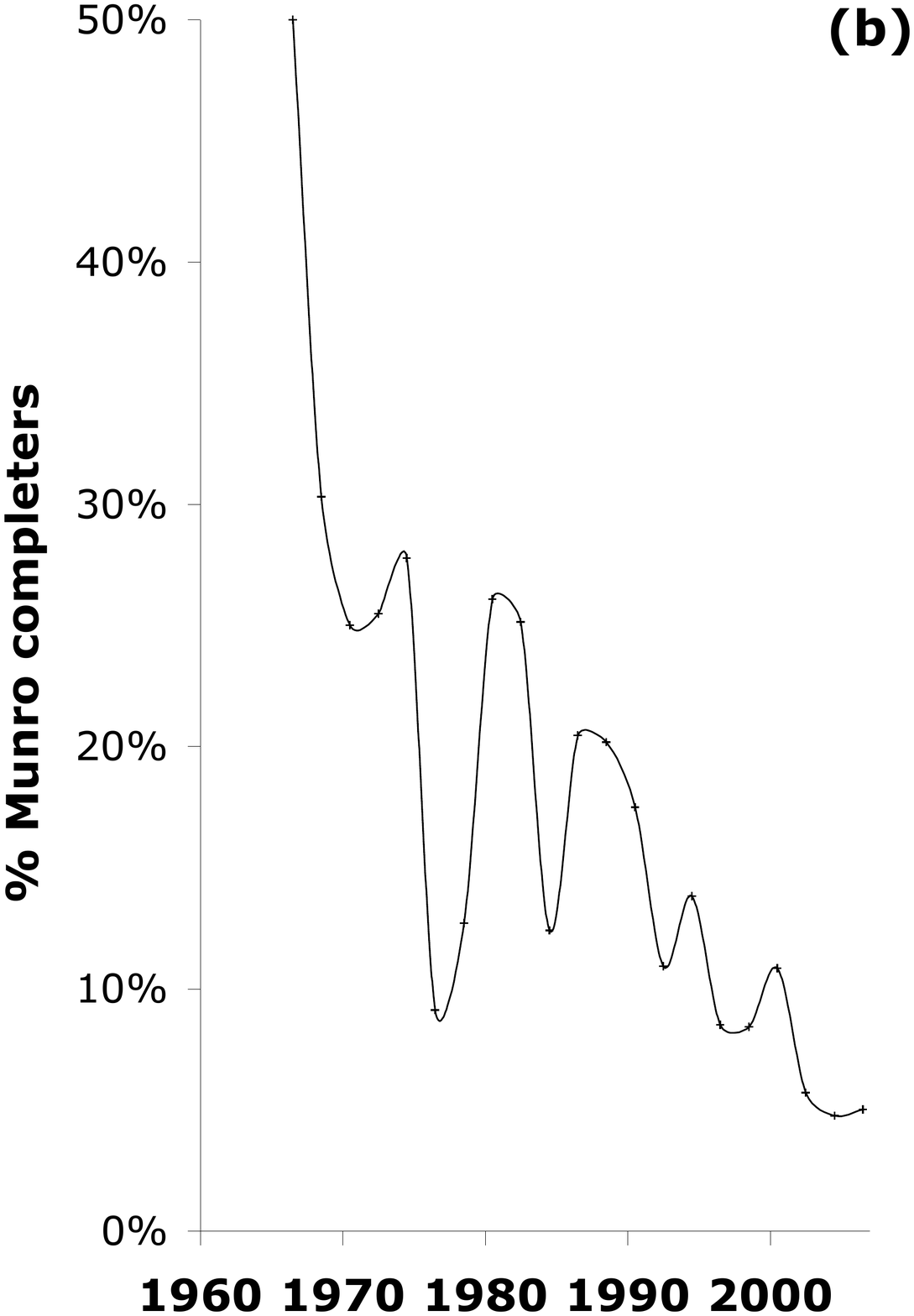}
\end{center}
\caption{\textbf{(a)} Tops completers each year, 1965-2007. \textbf{(b)} Tops completers as a percentage of Munros completers, 1965-2007, in two-year averaged intervals.}
\end{figure}
%
\section{Theoretical models: Fashions versus traditions}
To model a `fashion', we consider a model of selective copying among binary options, after the Bass diffusion model \cite{Bass_1969}. The Bass model assumes that the spread of a behaviour depends on random encounters between `potential' and `actual' adopters, with a fixed probability that any such encounter causes the `potential' to adopt. The model implies that people adopt a particular behaviour mainly from imitating others, with the probability of adoption at time $t$ is modelled as:
\begin{equation}
p(t) = (\mu + qF(t))(1-F(t))
\end{equation}
where the parameters $\mu$ and $q$ represent the degree of innovation and imitation, respectively. This well-known model assumes a binary decision, between whether or not to adopt a single behaviour.  The first half of $p(t)$ in eq. (1) models the imitation rate as proportional to the fraction $F(t)$ who have already adopted the behaviour, governed by the constant $q$, and the relative rate of independent discovery, governed by $\mu$. The second half of $p(t)$ forces the levelling off of the spread of behaviour as it approaches the full number of eventual adopters.  This is a standard aspect of logistic growth or `S' curve models, in which the rate of adoption decreases exponentially as the cumulative number of adopters asymptotically approaches its pre-determined maximum.  
\\ \indent Empirical studies on durable consumer goods suggest that the innovation and imitation rates typically differ by an order of magnitude, with innovation µ about 1\% or less and imitation $q$ from 10\% to more than 50\% \cite{Srinivasan_Mason_1986}. This means that most people adopt the behaviour upon witnessing it, whereas only a minority independently decides to adopt it. In both aspects, the adoption is made based on the perceived benefits of the behaviour over the previous alternative. 
\\ \indent  If $N$ is the total number of eventual adopters, the cumulative number of adopters in (1), $X_t$,  is given by \cite{Schmittlein_Mahajan_1982}: 
\begin{equation}
X(t)=NF(t)=N\frac{1-e^{-(\mu+q)t}}{1+\frac{q}{\mu}e^{-(\mu+q)t}}
\end{equation}
The number of new converts in period $t$ is $X_t-X_{t-1}$ or:
\begin{equation}
X_t=X_{t-1}=\mu N+(q-\mu)X_{t-1}-\frac{q}{N}X_{t-1}^2.
\end{equation}
After an initial growth period, the Bass model yields an exponential decay in the number of new converts, as the cumulative behaviour asymptotically approaches its maximum frequency, or saturation point. 
\\ \indent Following this pattern, the Tops demonstrate a marked decrease in the number of completers per year over the last decade (Figure 2). 
\\ \indent Over the whole data period, 1901-2007, the fit of the simple Bass model for the Tops is characterised by estimates of $\mu$, $q$ and $N$ which are robust with respect to the choice of sub-samples within the entire period. The estimates are carried out imposing the implied constraints on the coefficients in (3), using non-linear least squares regression (with the command `nlsÕ in the statistical package S-Plus). Again, detailed results are in the \textit{SJPE} paper.
\\ \indent 
A feature of the results is that the coefficient on $\mu$ is not significantly different from zero in any individual equation, while $q$ and $N$ are highly significant.
\\ \indent In contrast, the fashion model does not account for the Munros very well, mainly because the number of Munros completers per year is persistent after its decline in growth, rather than decaying (Figure 1). 
\\ \indent At first sight, the results obtained using the full Bass model (eq. 3) for the Munros appear very similar to those obtained with the Tops.  The results are set out in Table 1.
\begin{table}[htbp]
\caption{Estimations of simple Bass model for the Munros.}
\begin{center}
\begin{tabular}{lcccc}
\ Period&$\mu$&$q$&$N$\\\hline 1901-2007&-0.0001&0.143&6023\\1951-2007&-0.0003&0.144&5994\\1951-1995&-0.0014&0.173&3956\\1961-1995&-0.0002&0.186&3536\\1961-2007&-0.0003&0.145&5979
\end{tabular}
\end{center}
\end{table}
\\ \indent Again, the estimates of $\mu$ are not statistically significant from zero (except for 1961-1995 which is just significance at $p = 0.05$).  And the value of $q$ are similar to those obtained for the Tops.
\\ \indent This similarity extends to the fact that, over the entire sample period, the null hypothesis that the Tops and Munros follow the same statistical distribution\footnote[2]{Normalised by the respective sums, so that the annual rates are the percentage of the total 1901-2007 in each year.} is only rejected on a Kolmogorov-Smirnov test at the p-value= 0.0531, and for the 1951-2007 at 0.0639.  In other words, on the conventional criterion of significance (0.05)  the null hypothesis cannot be rejected, although the calculated values are of course close to the rejection level.
\\ \indent As noted above, however, there is an important difference between the Tops and Munros in that the latter have remained close to their peak level in recent years, whereas the former have fallen away sharply.  A reflection of this is that the saturation level $N$ is much more sensitive to the sample period in the Munros case (Table 1). 
\\ \indent Given that empirical estimates show that $\mu$ is in general not significantly different from zero, we can simplify eq. (3) to give:
\begin{equation}
X_t-X_{t-1}=qX_{t-1}-\frac{q}{N}X_{t-1}^2.
\end{equation}
Recognising this parameter $N$ as the deciding difference, the reason eq. (4) does fit the Munros would appear to be population flux.  In other words, the potential number of Munro completers is not time-invariant.
\\ \indent We modify the simplified model in eq. (4) to incorporate population flux into the number of new adopters per time interval. If the population is growing, then the maximum number of adopters $N_t$ increases accordingly.  Letting $a$ denote fractional rate of the growth in $N_t$ per year, such that $N_t= (1+a)N_{t-1}$, eq. (4) becomes:
\begin{equation}
X_t-X_{t-1}=qX_{t-1}-\frac{q}{(1+a)N_{t-1}}X_{t-1}^2.
\end{equation}
The first term on the right hand side is the same as in eq. (4), whereas the modified term reflects how a growing population introduces an additional number of potential adopters per time interval.  
\\ \indent This persistence is why we consider this a model of `traditions', as the activity is carried on through time and indeed passed on between generations (population renewing through time).
\\ \indent In terms of estimating (5), we initially create a new variable, $(1 + a)N_{t-1}$, setting $N$ equal to approximately 6000 in 2007, and estimating (5) for  $a =0.01, 0.02, 0.03, 0.04$ and 0.05.  We consider the period 1960-2007, since even with the Munros there were only 36 completions prior to that date.  By the end of 2007, the actual number of completers was 4019, so implicitly there were then around 2000 people consciously engaged in the process of bagging the Munros but who had not yet done so.  This, of course, is a fairly arbitrary but not unreasonable figure to use. 
\\ \indent The residual standard error of the equation for the various values of a are, respectively, 18.16, 17.80, 17.63, 17.93 and 18.38.  We then refined the gird search, as it were, and used values of a between 0.022, 0.024, \ldots 0.038.  The standard error of the equations is rather flat for values between 0.026 and 0.036.  The actual minimum is at $a = 0.028$, and we report results with this value.
\\ \indent The constraint in (5) that the coefficients on $X_{t-1}$  and $\frac{X_{t-1}^2}{(1+a)N_{t-1}}$ are the same in absolute value was tested and is not rejected by the data, the standard errors of the equation being, respectively, 17.43 and 17.59 without and with the constraint.
\begin{equation}
X_t-X_{t-1}=0.164\left(X_{t-1}-\frac{X_{t-1}^2}{(1+a)N_{t-1}}\right)
\end{equation}
 \begin{equation}                   (0.0037)
\end{equation}
Adj $R^2 = 0.977$, Skewness (1) =0.78, Kurtosis (1) = 2.55, LM(3) = 5.52
\vspace{5 mm} 
\\where $a = 0.028$, the figures in brackets is the standard error on the estimated coefficient (there is no constant term), Skewness and Kurtosis are chi-square variables each with 1 degree of freedom of the null hypothesis that the residuals exhibit no skewness and no kurtosis, and LM(3) is the Breusch-Godfrey test of the null hypothesis that the residuals exhibit no auto-correlation at lags 1 through 3, and is distributed as a chi-square variable with 3 degrees freedom.  The regression was carried out in STATA, which does not allow the Ramsey test for omitted variables to be carried out in regressions without a constant.  However, estimating the equation with a constant gives a calculated value for the Ramsey test, an $F$-distribution with (3,42) degrees of freedom of 1.15.
\\ \indent The estimated equation therefore appears to be well specified.  Over the 2000-2007 period, the annual average of actual completions was 203 compared to the predictions of the model of 213, but this is accounted for by the unusually low number of completions in 2001, just 149.
\\ \indent In contrast, the simplified Bass model in equation (4), omitting $\mu$, gives a good explanation for the Tops, again using data 1960-2007.  Imposing the value of $N$ from Table 1, we get
\begin{equation}
X_t-X_{t-1}=0.154\left(X_{t-1}-\frac{X_{t-1}^2}{498.6}\right)
\end{equation}
\begin{equation}                  (0.008)
\end{equation}
Adj $R^2 = 0.977$, Skewness (1) = 13.4, Kurtosis (1) = 1.49, LM(3) = 1.78
\vspace{5 mm} 
\\ The overall level of fit is less than with the preferred equation for the Munros, because the small numbers of the Tops completers exhibits more year-by-year random fluctuation, but it is still good.  In particular, the equation captures the reduction in Top completers towards the end of the sample period, the fitted values averaging 11.4 for 2000-2007 compared to the actual average of 13.  The 1990-1999 average is 18, peaking at 29 in 1991.
\\ \indent The residuals exhibit skewness, but the simple model is otherwise well-specified.  Again including a constant term, the Ramsey test statistic for omitted variables ($F(3,42)$) is calculated as 0.93.
%
%
\section{Conclusion}
We consider in this paper a market in which consumers select not simply on the basis of the perceived attributes of products, but in which their preferences might be modified by the behaviour of others.
\\ \indent The specific `market' is the consumption experience of ascending all the Munros, a list of 3000 foot peaks in the Scottish Highlands.  We examine also the completion of the list of Munro Tops, a difficult and time consuming extension of the Munros themselves.  Completions of either or both lists can be registered with the Scottish Mountaineering Council (SMC).
\\ \indent The SMC data show that the annual number of registered Munro completions levelled out in the late 1990s.  In contrast, the (much smaller) number of those completing the Tops as well as the Munros peaked in 1991 and has since fallen away markedly.  
\\ \indent An important feature of this data is that it embodies substantial changes over time in the population making choices at any given point.  
\\ \indent The classic Bass diffusion model, which incorporates the imitation of others as a part of the behavioural rules used by consumers in making choice.  We extend this model to be able to apply it to long-term case studies which involve successive cohorts of consumers.  The applications of the extension we make to both the Munros and Tops are illustrations of how the model can be used more generally in relevant contexts.
\bibliographystyle{unsrt}

\end{document}